\documentclass[preprint,12pt,3p]{elsarticle}

\usepackage{amssymb}
\usepackage{xcolor}

\journal{Methods}

\begin{document}

\begin{frontmatter}

\title{Shedding light on the dark matter of the\\ biomolecular structural universe:\\ Progress in RNA 3D structure prediction}

\author[label1]{Fabrizio Pucci}
\address[label1]{John von Neumann Institute for Computing, J\"ulich Supercomputer Centre,\\ Forschungszentrum J\"ulich, 52428 J\"ulich, Germany}
%address[label2]{3BIO-Computational Biology and Bioinformatics, Universit\'e Libre de Bruxelles, 1050 Brussels, Belgium}
\ead{f.pucci@fz-juelich.de}

%\fntext[label3]{I also want to inform about\ldots}
%\fntext[label4]{Small city}

\author[label1]{Alexander Schug\corref{cor1}}
\cortext[cor1]{corresponding author}
\ead{al.schug@fz-juelich.de}

\begin{abstract}
Structured RNA plays many functionally relevant roles in molecular life. Structural information, while required to understand the functional cycles in detail, is challenging to gather. Computational methods promise to complement experimental efforts by predicting three-dimensional RNA models. Here, we provide a concise view of the state of the art methodologies with a focus on the strengths and the weaknesses of the different approaches. Furthermore, we analyzed the recent developments regarding the use of coevolutionary information and how it can boost the prediction performances. We finally discuss some open perspectives and challenges for the near future in the RNA structural stability field.    
\end{abstract}


\end{frontmatter}

% \linenumbers


\section{Introduction}
In the last two decades growing attention has been dedicated to the understanding of RNA. As for proteins, RNA structure and function are closely tied and play a determining role in many biomolecular processes such as the splicing process, transcriptional and translational machineries, and RNA localization and decay \cite{ReviewA}. Despite this importance, the number of available experimental RNA structures at an atomic level stored in public databases such as the Protein Data Bank (PDB) \cite{PDB} or the Nucleic Acids Database (NDB) \cite{NDB} remains limited due to challenging experimental problems related to the preparation and/or crystallization of RNAs that are usually more flexible and dynamic with respect to proteins \cite{ReviewB}. Currently, more than 90\% of structures stored in the PDB database \cite{PDB} are proteins, while less than 5\% of the human genome encodes for proteins. This discrepancy has stirred the curiosity of scientists and lead to the remaining 95\% of the human genome sometimes being referred to as the dark matter of the genome \cite{DarkMatter1, DarkMatter2}. 

To overcome the lack of structurally resolved RNA, computational methods have complemented experimental efforts to get more insight into how RNA structure and dynamics determine its functions \cite{Firstpaper,Join,Pappa}. Significant efforts have been devoted to the construction of methods to predict the RNA secondary structure mainly employing thermodynamics-based models \cite{2DReview}. These methods have been recently achieved significant improvements by the incorporation of auxiliary structural information from high-throughput chemical probing technologies \cite{Join1,Join2}.

However, even if the knowledge of the RNA secondary structure provides important information, it is not sufficient to fully explain RNA function or interactions with other biomolecules \cite{Computational3D}.  During the last years lot of attention has been focused on the construction of RNA 3D structure prediction tools of increasing accuracy and speed \cite{CompModelingI, Adamiak, ModeRNA, Bujnicki, Chen, Das, iFold, 3dRNA, Baker, Altman, Alex, MCFOLD, ReviewI}. 

In this review we provide a concise overview of these methodologies, present their strengths and limitations, and highlight the open challenges in RNA structure prediction. We will particularly underline the recent development related to the use of coevolutionary information to improve the accuracy of the RNA 3D structure prediction methods. The structural information remains, however, static and provided one piece to the puzzle of RNA function. Another important component is the dynamics of RNA, for example while undergoing large conformational rearrangements \cite{RNA_Dyn_I, RNA_Dyn_II}, which is exhaustively covered in the excellent review \cite{ReviewBIG}.  

\section{From the RNA sequence to its 3D structure}

The basic unit of RNA is the nucleotide that is formed by planar aromatic rings linked to a ribose unit that in turn is attached to a phospate group (see fig 1). The sequence of the different constituent nucleotides (adenine, guanine, cytosine, uracil) of a given RNA molecule is defined as its \textbf{primary structure}.      

Nucleotides typically complement each other by forming the canonical base pairs A-U and C-G, which maximizes inter-nucleotide hydrogen bonding. This leads to short chains of nucleotides folding in anti parallel double helices. The nucleotides that do not form Watson-Crick base pairs can remain unpaired or establish less stable non-canonical base pairs forming internal and bulge loops, hairpins and junctions. The \textbf{secondary structure} is thus essentially defined as the set of base pairs occurring in the RNA molecules.      

The \textbf{tertiary structure} is the complete set of three-dimensional coordinates of all atoms of the RNA structure. This includes formation of a plethora of tertiary motifs such as pseudoknots, stacking of helices, multiple base pairing, ribose zipper and loop-loop interactions that determine the molecular shape in the physical space.

An accurate computational prediction of the RNA tertiary structure starting from its sequence is particularly challenging as the 3D structure depends not only on the sequence but also on the environmental conditions such as the ion concentrations and temperature. % Citation
% Should we keep this line? It complicates things?

\begin{figure}[!h]%figure2
	\begin{center}
		\includegraphics[width=17cm]{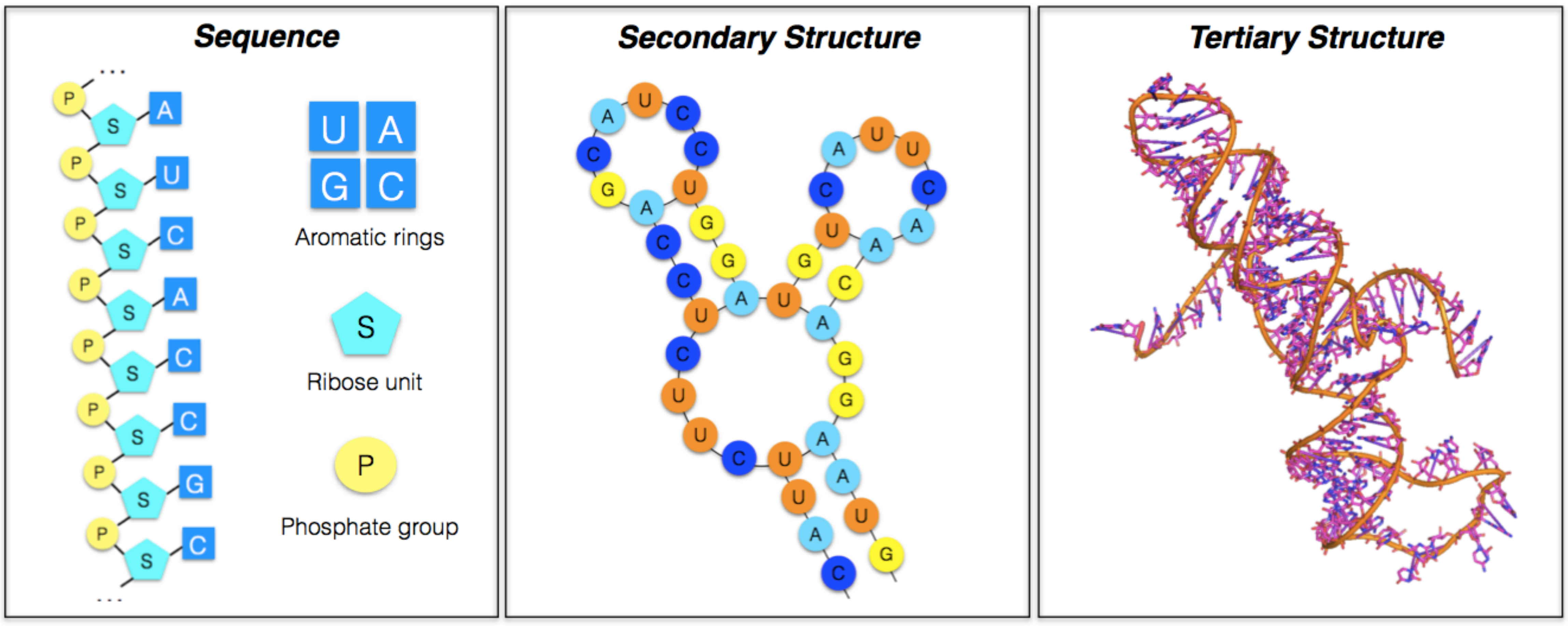}
	\end{center}
	\caption{From primary (sequence), to secondary and to tertiary RNA structures.}
\end{figure}

\section{Computation modeling of RNA 3D structure}

Here, we review and compare some widely known methods for the prediction of the three-dimensional structure of RNA. The available approaches can be roughly divided in three different types: fragement-based, physics-based and comparative modeling. To compare the state of the art prediction methods and assess their performance, a blind experiment for the RNA 3D structure prediction has been established in the last years \cite{RNA-PuzzleI,RNA-PuzzleII,RNA-PuzzleIII} with the last round focused on the challenging prediction of six RNA structures of riboswitches and ribozymes \cite{RNA-PuzzleIII}.

\subsection{Fragment-based homology methods}

The main idea behind this approach is to assemble the 3D prediction of target molecules using small fragments from libraries with similar sub-sequences. The theoretical justification of such a procedure comes from assuming that the distribution of the different conformations observed in known RNA structures for given fragment sequences is a good approximation for the conformation of similar or identical sub-sequences. 

The basics steps of these methods consist first in the fragmentation of the secondary structure used as input. As a second step a search algorithm is employed to match these elements from fragment libraries constructed from databases of known RNA structures. Finally, all the elements are assembled together using different algorithms (see below) and, usually, a final refinement stage using atomic force field or coarse-grained potentials is  performed.

One advantage of these methods is their computational efficiency as the fragmentation assembly drastically reduces the conformational search space. As the structural diversity of the fragment library directly limits the accuracy of the composed assembly, good results require a large and diverse library as well as a good scoring function. Here, we list methods belonging to this class and some of their characteristics.

\begin{itemize}

\item \textbf{RNAComposer} \cite{Adamiak}: after the fragmentation step, the predicted secondary structure elements (stem, loops and single strands) constitute the input pattern for a search in the FRABASE 3D fragment data-set developed by the authors. From the matched elements a 3D structure  is constructed by first superimposing and then merging them. Finally, an energy minimization is performed in the CHARMM force field \cite{CHARMM}.

\item \textbf{Vfold3D} \cite{Chen} uses a coarse-grained representation of the RNA. First, it utilizes VFold2D, a free energy-based model, to predict the secondary structure from which it extracts motifs (helices, hairpin loops, internal loops,...). From these motifs it searches the best template in the VFoldMTF database. After assembling the 3D structural motifs and the addition of all atoms to the coarse-grained structure (according to the template) it performs an all-atom structure refinement. 

\item \textbf{3dRNA} \cite{3dRNA} uses a two-steps procedure where first the smallest secondary elements (SSEs) are assembled in hairpins and duplexes one by one following the 5' to 3' end direction. Then, these structures are further assembled into a complete tertiary structure by selecting the junction component from a junction database. Finally, to assure the chain connectivity, the  assembled model is energy minimized in the AMBER 98 force field \cite{AMBER}.

\item \textbf{FARNA} \cite{Baker} (Fragment Assembly of RNA) also uses a coarse-grained representation of the RNA structure and a fragment assembly strategy employing a Monte Carlo process that is guided by a low-resolution knowledge-based energy function. The authors developed knowledge-based base-pairing and base-stacking potentials to which they add several other terms such as the penalty for steric clashes. The structural model undergoes a second refining step in an all-atom potential to improve the accuracy and to better discriminate competing structural models. The two-step protocol is called \textbf{FARFAR} (Fragment Assembly of RNA with Full Atom Refinement) and is part of the ROSETTA package. 

\item \textbf{MC-Fold/MC-Sym} \cite{MCFOLD} this pipeline uses the combination of small motifs called nuclotide cyclic motifs (NCMs). The NCM-3D fragments are assigned to the given sequence by choosing the structure with higher probability of occurencies. Then the structural NCMs are concatenated using a Las Vegas probabilistic algorithm. 

\end{itemize}

\subsection{Physics-based methods}

In contrast to the previous methods, the physics-based models do not use template structures in the assembly of RNA fragment/motifs but derive and parameterize energy functions depending on specific conformations, similar to approaches applied for proteins \cite{schug2003,schug2005}. These methods can be further separated in \emph{ab-initio} approaches or knowledge-based approaches. In the latter methods, the energetic functions are derived using the inverse Boltzmann law from the probability of occurrences of certain sequence-structure elements in a dataset of known structures.
In constrast, the \emph{ab-initio} methods are based on force fields adopting usually harmonic potentials for bond lengths and angles, Lennard-Jones potentials for Van
der Waals interactions, and electrostatic potentials that get reparameterized based on RNA structure and thermodynamics data.

Such energetic functions are then used in Molecular Dynamics (MD) simulations or Monte Carlo (MC) minimization often associated with enhanced sampling techniques such as temperature replica exchange or discrete molecular dynamics simulations in which the energy function is substituted with discrete step function potentials that drastically reduce the computational cost of the method. 

The strength of the physics-based methods is that they are applicable to sequences with no known similar sequences or even sub-sequences. Their disadvantage is their need to explore a large conformational search space which increases computational demands and decreases their computational efficiency in comparison to fragment-based methods.

Here in the following the list of the computational tools that use this approaches.

\begin{itemize}

\item \textbf{iFoldRNA} \cite{iFold} uses a simplified "three-bead per nucleotide" representation of the RNA structure, and it is based on a replica-exchange discrete molecular dynamic (DMD) simulations protocol to span the conformational space. DMD incorporates base-pairing and base-stacking interactions into an energy function where in addition an entropic estimation of the loop formation is also considered.

\item \textbf{NAST} \cite{Altman} uses coarse-grained representation of RNA considering one quasi-atom per RNA nucleotide base. A simplified knowledge-based energy function, derived from the observed RNA geometries at the nucleotide level, is used to predict the target structure by global energy minimization. NAST necessitates as input the (known or predicted) secondary structure information and accepts also tertiary contacts to guide the folding.

\item \textbf{SimRNA} \cite{Bujnicki} uses a coarse-grained representation of the RNA structures  reducing the number of explicitly represented atoms per residue from about thirty to only five. It is based on dedicated RNA statistical potentials to compute the structure free energy and identify the native structure via Monte Carlo sampling.
\end{itemize}

\subsection{Comparative homology-based modeling}

Another type of methods uses homology modeling approaches by identifying structurally related template and  geometrically aligning residues from the target onto corresponding residues in the template. Examples of this type of methods are \textbf{RNABuilder} \cite{ComparativeI} and \textbf{ModeRNA} \cite{ModeRNA}. The latter makes also extensively use of the evolutionary information by using multiple RNA sequence alignments to better reveal patterns of conservation that improve the accuracy of the prediction starting from the 3D template.

In order to improve the accuracy of homology-based methods it has been shown that the addition of multiple templates can be successfully employed. Another characteristic that is common to this type of methods is that they model (short) regions with no template by employing fragment-based insertion approaches. Finally the methods perform usually a geometry optimization using a force field in order to obtain physically reasonable conformations. The RNABuilder method for example uses a multi-resolution approach that handles at different level of resolution the forces, rigidifying certain bonds, residues or molecules part while keeping flexible the others.

The major drawback of this class of methods resides in the difficulty of having the template structure for the given sequence and an informative multiple sequence alignment. Indeed for the RNA structure templates there is the limitation of the number of structure deposited in the widely known database \cite{NDB}. Regarding the alignments one can use those available for many RNA families in the Rfam database \cite{Rfam} or perform alignment via commonly used multiple RNA sequence alignment packages such as R-Coffe \cite{RCoffe}, Muscle \cite{Muscle} or Infernal \cite{Infernal}.

The strength of these methods is their high accuracy with modest computational costs when good structural templates can be found. Their performance drops in the absence of such templates.

\subsection{Performance assessment and RNA-Puzzles prediction}

Most of the analyzed RNA structure prediction methods participated in the RNA-Puzzle competitions \cite{RNA-PuzzleI,RNA-PuzzleII,RNA-PuzzleIII} in which a set of experimentally resolved RNA 3D structures had to be blindly predicted. To assess the performance of the predictors and rank the models, different metrics have been used such as the root mean square deviation (RMSD) between the predicted the experimental crystal structures that gives a more global information about the model's accuracy or the deformation index and the complete deformation profile matrix that instead capture the "local" accuracy at the nucleotide interaction level. 

In the first RNA-Puzzle round \cite{RNA-PuzzleI} in addition to two simple small targets that were relatively well predicted, the more challenging riboswitch structure were not accurately reproduced with a mean RMSD accuracy of about 15\, \AA. Moreover while most methods achieve good performance on Watson-Crick base pairs, non-Watson-Crick interactions remain difficult to predict and clash score remains generally quite high.

In the second RNA-Puzzle round \cite{RNA-PuzzleII} the best RMSDs for a long nucleotide sequence range between 6.8 and 11.7\, \AA\, indicating a global improvement of methods' performance. A substantial amelioration for non-Watson-Crick interactions prediction is also observed.

Finally in the last RNA-Puzzle competition \cite{RNA-PuzzleIII}, the predictions achieved a consistently high level of accuracy especially when a high-homology template can be identified. For example in the case of the SAM-I riboswitch aptamer prediction that has as template (PDB code 3QIR), the average RMSD over all predicted models is about 4.3 \AA , with a standard deviation of less then 2 \AA .

Unfortunately, when the homology with the template is not high enough, the accuracy of methods is still not satisfactory and depends on the length of the RNA sequence. Small RNA sequences can be predicted with good accuracy as exemplified in the case of the ZTP riboswitch predicted with an averaged RMSD of about 6 \AA\, and as also shown in the previous RNA-Puzzle round. For long sequences such as the \emph{ydaO} riboswitch no method is capable of reliably predicting the native three dimensional conformation with an average RMSD of about 16 \AA. 

In order to improve the structure prediction of these challenging targets, there is a need for new and more performing algorithms. In the next section we will thus present recent progress in this direction and more in detail we will show how the coevolutionary information can been used to improve significantly the methods' accuracy.

\begin{figure}[!h]%figure2
\begin{center}
\includegraphics[width=16.5cm]{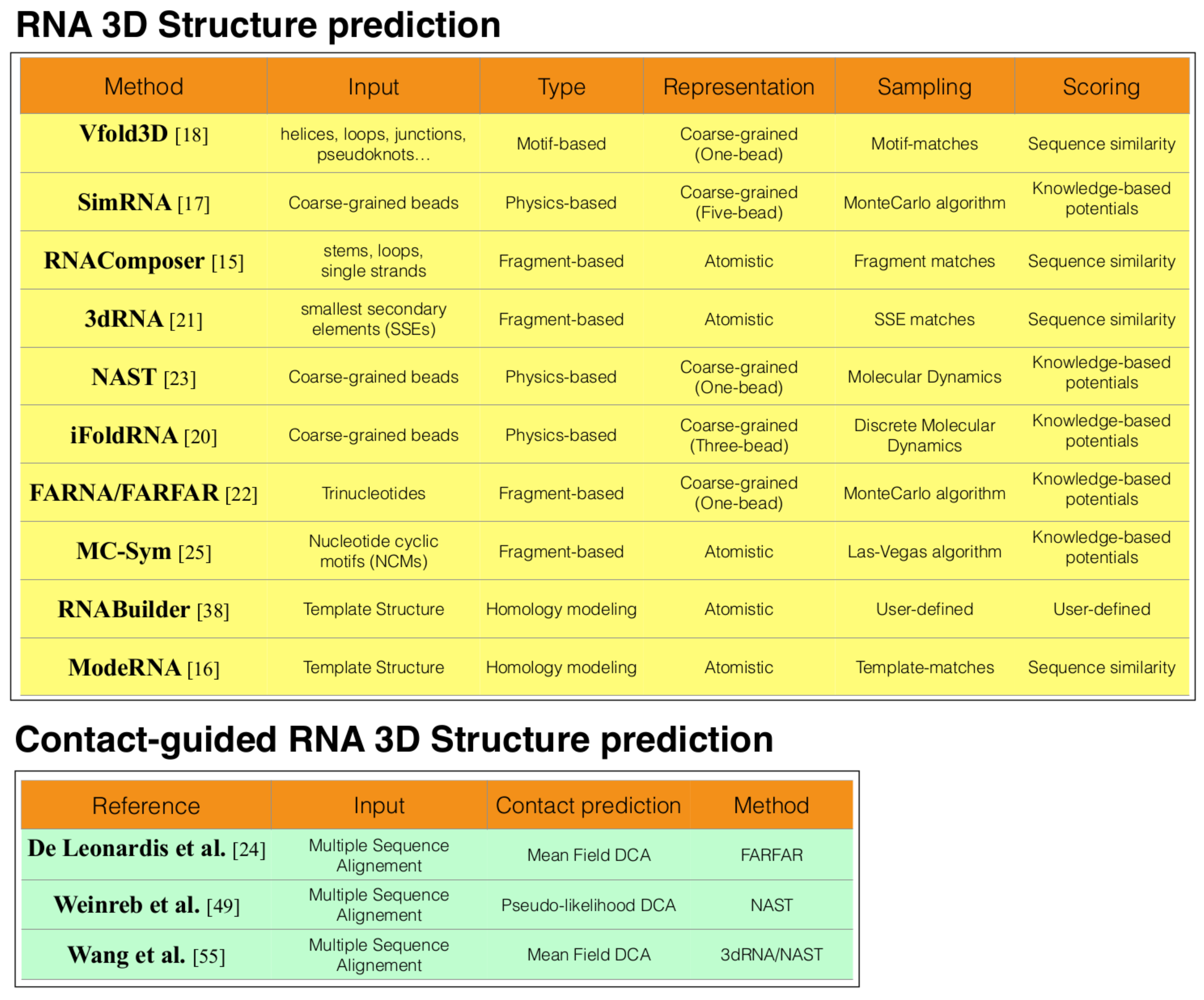}
\end{center}
\caption{3D RNA structure prediction methods and their principal characteristics}
\end{figure}

\section{Including evolutionary information to improve 3D structure prediction}

\subsection{Residue co-evolution and contact prediction}

A significant amount of data obtained from high-throughput sequencing technologies provides us an invaluable source of evolutionary information that can be used in order to improve the protein \cite{Weigt2009, Schug2009, Marks, Weigt, Dago2012} and RNA structure prediction  \cite{Alex, MarksIII}. The basic idea behind these approaches is tracing co-variation of amino acid or nucleic acid pairs in proteins and RNA belonging to homologous families. Such co-variation indicates structural proximity of the involved residues and is hence related to biomolecular structural and stability properties. Compensatory mutations occur when a mutation with a detrimental effect at a given site, interact with a secondary mutations at another site to restore the molecular fitness \cite{Dimitri} thus indicating the tendency of co-evolving residues to represent physical interactions that are important for the stability and function of biomolecules.

In the last decade many statistical methods have been developed to identify co-evolving residue pairs in a multiple sequence alignment (MSA) \cite{Alfonso}. One can assume that such correlation occurs due to the spatial proximity of the two residues even if it can also arise from indirect effects related to the transitivity of the interaction between pairs and tertiary residues. 

The use of the statistical methods such as maximum entropy models (MEM) or direct coupling analysis \cite{Alex, Weigt2009, Dago2012, Weigt} allows to unravel the transitive effects in the network of constrained residue-residue interactions and thus they give more efficient and robust contact-prediction. Using these statistical-based approaches one can detect long-range tertiary contacts from sequence covariation whose prediction difficulty has been one of the main limitation to the advancement of the computational RNA 3D structure prediction methods.

\begin{figure}[!h]%figure2
\begin{center}
\includegraphics[width=16cm]{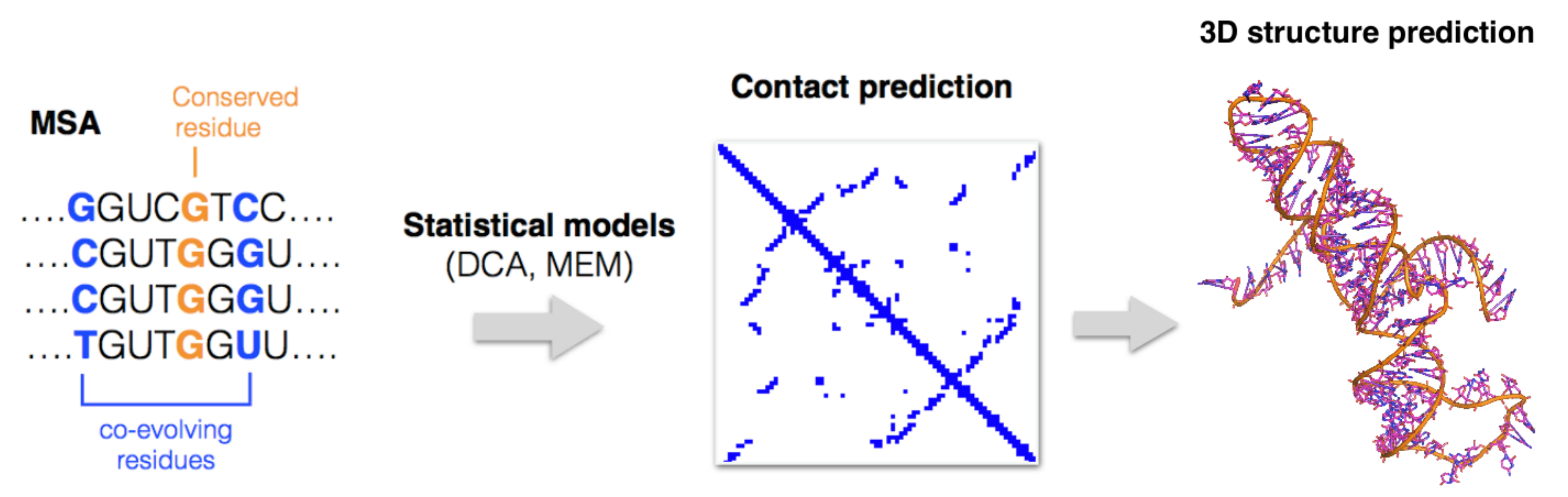}
\end{center}
\caption{Statistical-based contact prediction from coevolutionary data improved the 3D RNA structure prediction}
\end{figure}

\subsection{Direct coupling analysis (DCA)}

The basic assumption of this method is to associate 
the probability of observation $P(\sigma)$ of a given sequence $\sigma$ = ($a_1$, $a_2$ ...$a_L$) of length $L$ in a MSA to the Hamiltonian energetic function $H(\sigma)$ using the Boltzmann law

\begin{equation}
P(\sigma)=\frac{1}{\mathcal{Z}}e^{- \beta H(\sigma)}  
%P(\sigma)=\frac{1}{\mathcal{Z}} \text{exp}\legth(- \beta H(\sigma)\rigth)  
\end{equation}

\noindent
where $\beta$ is the temperature and $\mathcal{Z}$ is the partition function of the system, and where the Hamiltonian is assumed to have the following simplified form 

\begin{equation}
H(\sigma) = -\sum_i^L h_i (a_i) -\sum_i^{L-1}\sum_{j=i+1}^{L} J_{ij} (a_i,a_j)      
\end{equation}

consisting only of single site terms, \emph{i.e.} $h_i (a_i)$, and residue pair interactions $J_{ij} (a_i,a_j)$. These parameters can be inferred from the MSA using a plethora of different approaches. For example in \cite{MarksIII} a pseudo-maximum likelihood (pmlDCA) approximation have been employed, while a computational intensive message-passing algorithm (mpDCA) is used in \cite{Weigt2009} and a more efficient  mean field algorithm (mfDCA) in \cite{Weigt}. The list of other type of popular algorithm used in the inverse inference step can be found in \cite{Alex}.

There are also pitfalls. Frequently, some species are over-represented in the MSA, e.g. because of their medical importance or the ease of handling them experimentally. Thus, these sequences need to be re-weighted. In addition, the quality of the MSA such as the proper placement of gap regions influences contact prediction accuracy. This loss of contact prediction precision directly leads to a decreased quality of 3D prediction. Another drawback is that the DCA prediction of the tertiary contacts is far from being perfect with only a modest overall true positive (TP) predicted contacts; it should be noted, however, that only relatively few (O(10)) higher ranking pairs that show higher TP rate are already sufficient to boost the performance of the structural modeling. Still, these methods significantly boost performance without too much computational effort.

\subsection{Contact guided 3D RNA-structure prediction}

While the use of coevolutionary data has been already fruitfully applied to protein structure determination during the last decade \cite{Marks,MarksI,Weigt2009,Schug2009,Dago2012,Weigt,Baker222,Sulkowska,Jones}, the contact guided prediction of the three-dimensional RNA structure is relatively new. Indeed, the previous mutual information (local) approach to the extraction of coevolution signals from MSA was not sufficiently accurate \cite{West} to provide reliable tertiary contact predictions.  

Recent investigations \cite{Alex,MarksIII,Wang} instead show that the use of a global approach to extract the top-ranked site-pairs with stronger co-evolutionary signals can be efficiently employed as distance constraints in modeling tools. 

In \cite{Alex} the authors show that in the prediction of the structure of six representative riboswitches with the Rosetta-based method FARFAR, the use of predicted tertiary contacts by mfDCA improves the RMSD in average by about 30\% with respect to the case in which only secondary structure information (SSI) is provided. In figure 4 we report this explicit comparison for all the six structure considered.

\begin{figure}[!h]%figure2
\begin{center}
\includegraphics[width=15.5cm]{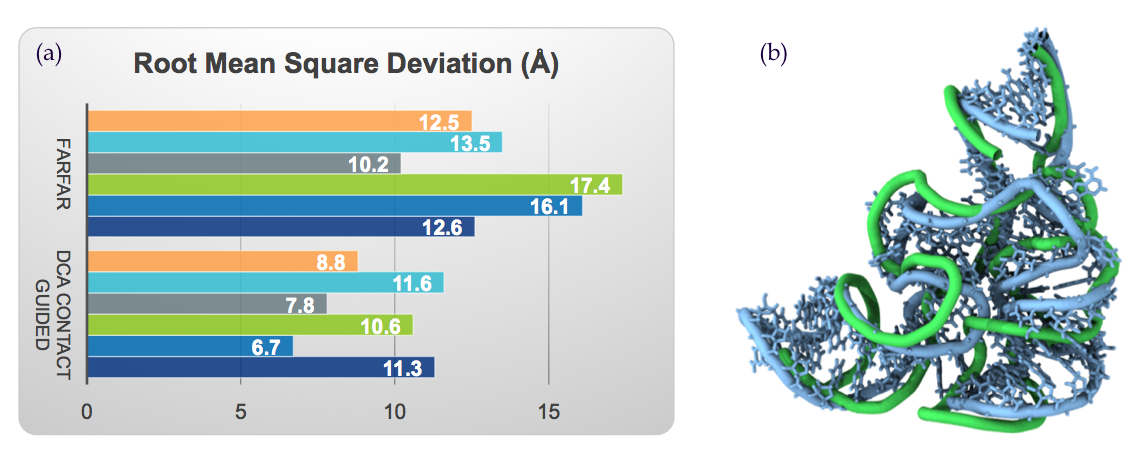}
\end{center}
\caption{(a) DCA contact-guided RNA structure prediction improvement with respect to the state of the art (Rosetta-based) method for the six riboswitches from \cite{Alex}. (b) Overlay of the  DCA-contact guided predicted (blue) and the experimental structure (green) for the thiamine pyrophosphate-specific (TPP) riboswitch (PDB code 2gdi). In the prediction, the first 100 top contacts as computed via mean field DCA from the MSA of the RF00059 family have been used as constraints in the FARFAR method.}
\end{figure}

These results have been confirmed in \cite{Marks} where the authors show a significant improvement of prediction quality when the evolutionary based contact prediction computed via the pmlDCA approach. In this work, contacts are used as spatial constraints in the NAST coarse-grained structure prediction method. A further confirmation in \cite{Wang} highlights that prediction RMSDs for the same structures as analyzed in \cite{Alex,Marks} are lowered by about 30\% when using the tertiary contacts predicted via mfDCA in the 3dRNA method  \cite{3dRNA} compared to not using such tertiary contact constraints. \cite{Marks} and \cite{Zhou} also demonstrated how DCA-based methods show good accuracy in the prediction of intermolecular RNA-protein contacts.

\section{Future challenges and outlook}

Even if in the last decade tremendous advances has been achieved in RNA structure prediction, its accuracy still is not as high as for protein structure prediction.  Moreover there are open and intriguing challenges in the field that will hopefully be tackled in the close future:

\begin{itemize}
\item The role played by the environmental conditions such as ions that strongly influence the RNA structure has to be fully investigated and clarified \cite{IONSI,IONSII,IONSIII}. Since \emph{in vivo} RNA can adapt different conformations with respect to \emph{in vitro} ones, this will be also important to understand such differences and give important information for RNA biology.  

\item In the next years, thanks to the advancement of the next generation sequencing technologies, the amount of sequence information will continue to increase exponentially. Currently coevolutionary methods focus on the prediction of two-site interactions (contacts), but this increased amount of information promises to also allow to predict higher order correlations that could further boost structure prediction methods.

\item Further improvements of RNA force fields will continue to increase the accuracy of predictions. These can help to better understand the role of the different RNA conformations, their stability and to gain new insights about the RNA structural dynamics. 

\item Combining structure prediction methods or simulations with experimental data such as Selective 2′-hydroxyl acylation analyzed by primer extension (SHAPE) \cite{Kirmizialtin2015}, Fluorescence Resonance Energy Transfer (FRET) \cite{Reinartz2018} or small angle X-Ray scattering (SAXS) \cite{JPCI, Weiel2019, JMB} will allow to probe RNA structures where a single method fails \cite{COMPMEETEXP}.

\item Inter-molecular protein interactions and contacts can be predicted via DCA and related methods \cite{szurmant2018}. This could be transferred to RNA.

\item Finally, it becomes more and more clear that base modifications such as the methylation or deamination play an important role in RNA biology by modifying the structure as well as the function of RNA. It could be thus of great interest in the next future to address and investigate these (epi)transcriptomics data to better understand all biological processes in which the RNA is involved.    

\end{itemize}

\end{document}